\documentclass[prl,twocolumn,showpacs,amsmath,superscriptaddress,psfig,amssymb]{revtex4}

\usepackage{graphicx}
\usepackage{dcolumn}
\usepackage{bm}
\usepackage{multirow}

\begin{document}


\title{Superconductivity and Disorder effect in TlNi$_2$Se$_{2-x}$S$_x$}

\author{Hangdong Wang}
\affiliation {Hangzhou Key Laboratory of Quantum Matter, Department of Physics, Hangzhou Normal University, Hangzhou 310036, China}
\affiliation {Department of Physics, Zhejiang University, Hangzhou 310027, China}

\author{Qianhui Mao}
\affiliation {Department of Physics, Zhejiang University, Hangzhou 310027, China}

\author{Huimin Chen}
\affiliation {Hangzhou Key Laboratory of Quantum Matter, Department of Physics, Hangzhou Normal University, Hangzhou 310036, China}

\author{Chiheng Dong}
\affiliation {Department of Physics, Zhejiang University, Hangzhou 310027, China}

\author{Rajwali Khan}
\affiliation {Department of Physics, Zhejiang University, Hangzhou 310027, China}

\author{Jinhu Yang}
\affiliation {Hangzhou Key Laboratory of Quantum Matter, Department of Physics, Hangzhou Normal University, Hangzhou 310036, China}

\author{Bin Chen}
\affiliation {Hangzhou Key Laboratory of Quantum Matter, Department of Physics, Hangzhou Normal University, Hangzhou 310036, China}

\author{Minghu Fang}
\email{mhfang@zju.edu.cn}
\affiliation {Department of Physics, Zhejiang University, Hangzhou 310027, China}

\date{\today}

\begin{abstract}
\noindent After our first discovery of the multi-band superconductivity (SC) in TlNi$_2$Se$_2$ crystal, we grew successfully a series of TlNi$_2$Se$_{2-x}$S$_x$ (0.0 $\leq$ x $\leq$ 2.0) single crystals. The measurements of resistivity, specific heat and susceptibility were carried out for them. Superconductivity with \textit{T}$_{C}$=2.3K was first observed in TlNi$_2$S$_2$ crystal, which also appears to involve heavy electrons with an effective mass m*=13$\thicksim$25 m$_{b}$, as inferred from the normal state electronic specific heat and the upper critical field, \textit{H}$_{C2}$(T), respectively. It was found that bulk superconductivity and heavy electron behavior is preserved in all the TlNi$_2$Se$_{2-x}$S$_x$ samples. In the mixed state, a novel change of the field dependence of the residual specific heat coefficient, $\gamma$$_{N}$(H), occurs in TlNi$_2$Se$_{2-x}$S$_x$, with increasing of the S content. We also found that the $T_{C}$ value changes with the disorder degree induced by the partial substitution of S for Se, characterized by the residual resistivity ratio (RRR). Thus, TlNi$_2$Se$_{2-x}$S$_x$ system provides a platform to study the effect of disorder on the multi-band superconductivity.
\end{abstract}

\pacs{74.70.Xa; 74.70.Tx; 74.25.Op; 71.27.+a}

\maketitle

The layered compounds with ThCr$_2$Si$_2$-type structure (space group \textit{I4/mmm}) exhibit versatile physical properties, including antiferromagnetic (AFM) ground state in BaFe$_2$As$_2$ \cite{ChenXH2009}, ferromagnetic (FM) ordering in (K,Rb)Co$_2$Se$_2$ \cite{YangJH2013}, Fe-based superconductivity (SC) with $T_C$ =30-50K in both (Ba,K)Fe$_2$As$_2$ \cite{Rotter2008} and (Tl,K,Rb)Fe$_x$Se$_2$ systems \cite{Fang2011,Wang2011}, as well as the heavy fermion SC in CeCu$_2$Si$_2$ \cite{Steglich1979}. Especially, SC emerges in both the Ni-pnictide compounds, such as BaNi$_2$As$_2$ ($T_C$ = 0.7K) \cite{Ronning2008}, SrNi$_2$P$_2$ ($T_C$ = 1.4K) \cite{Ronning2009}, in which the electron effective mass is not enhanced so much, and the Ni-chalcogenide compounds, such as KNi$_2$Se$_2$ ($T_C$ = 0.8K) \cite{Neilson2012-1}, KNi$_2$S$_2$ ($T_C$ = 0.46K) \cite{Neilson2013} and TlNi$_2$Se$_2$ ($T_C$ = 3.7K) \cite{WangHD2013}, in which SC appears to involve heavy electrons with an effective mass $m^*$=14-20 $m_{e}$. Due to Ni ions in the Ni-chalcogenide compounds having a mixed valence Ni$^{1.5+}$, it was suggested that the formation of a heavy-fermion state at low temperatures is driven by the hybridization of localized charges with conduction electrons, and the coherent state competes with a charge-fluctuating state \cite{Murray2012}. Thus it rises a question whether the SC close to a charge-fluctuation state in this system is unconventional or conventional.

TlNi$_2$S$_2$ crystalizes in a tetragonal ThCr$_2$Si$_2$-type structure (space group $I4/mmm$), as the same as that of TlNi$_2$Se$_2$. It can also be considered as one of a non-magnetic analogue of Fe-chalcogenide superconductor, \textit{i.e.} TlFe$_x$Se$_2$ compounds \cite{Fang2011,Wang2011}, recently discovered by us, but no superconductivity emerges in TlFe$_x$S$_2$ compounds with Fe vacancies. TlNi$_2$S$_2$ compound is also a Pauli paramagnetic metal, similar to that in TlNi$_2$Se$_2$ compound, reported first by A.R. Newmark \cite{Newmark 1989}, who did not observe any superconducting transition above 2 K. After our discovery of a multi-band SC in TlNi$_2$Se$_2$ crystal, we grew successfully a series of TlNi$_2$Se$_{2-x}$S$_x$ (0.0 $\leq$ \textit{x} $\leq$2.0) single crystals and rechecked their structure and physical properties. SC with \textit{T}$_C$=2.3 K has been observed in TlNi$_2$S$_2$, and appears also to involve heavy electrons with an effective mass $m^*$=14$\sim$20 $m_e$, although its \textit{T}$_C$ is a little lower than that in TlNi$_2$Se$_2$ ($T_C$=3.7 K). With the isovalent substitution of Se by S in TlNi$_2$Se$_{2-x}$S$_x$ crystals, the evolution of the superconducting properties were studied systematically. It was found that bulk superconductivity and heavy electron behavior preserves in all the TlNi$_2$Se$_{2-x}$S$_x$ compounds, which was confirmed by the results of resistivity, susceptibility and specific heat, respectively. It was found that, with increasing the S content \emph{x}, the magnetic field dependence of the residual specific heat coefficient at T = 0K in the mixed state, $\gamma$$_{N}$(\textit{H}), changes from a $\propto$$H$$^{0.5}$ observed in TlNi$_2$Se$_2$, to a linear relation, consistent with the prediction for a conventional \emph{s}-wave superconductor. Another, it was found that the \textit{T}$_C$ value in the TlNi$_2$Se$_{2-x}$S$_x$ crystals is related to the disorder, characterized by the RRR, which is induced by the partial substitution of S for Se.

Single crystals of TlNi$_{2}$Se$_{2-x}$S$_{x}$ (0.0 $\leq$ \textit{x} $\leq$2.0) were grown using a self-flux method. A mixture with ratio of Tl:NiSe:NiS = 1:(2-x):x was placed in an alumina crucible, sealed in an evacuated quartz tube. The mixture was heated at 950$^o$C for 12h, then slowly cooled to 650$^o$C at a rate of 6$^o$C/h, followed by furnace cooling. Single crystals with a typical dimension of 3$\times$3$\times$0.2 mm$^{3}$ were mechanically isolated from the flux, as shown in the left of Fig.1 for TlNi$_2$S$_2$ single crystal. The structure of single crystals was characterized by the X-ray diffraction (XRD). Figure 1 show the XRD pattern at room temperature for TlNi$_2$Se$_{2-x}$S$_x$ (\textit{x} = 0.0, 1.0 and 2.0) powder obtained by grinding pieces of crystals, other crystals have a similar XRD pattern (does not show here), which confirm that all the crystals of TlNi$_2$Se$_{2-x}$S$_x$ (0.0 $\leq$ \textit{x} $\leq$ 2.0) have the ThCr$_2$Si$_2$-type structure. With increasing of the S content, $\textit{x}$, the lattice parameters, \textit{a} and \textit{c} values decrease monotonously, as shown in the right inset of Fig.1, which is consistent with that the ionic radius of S$^{-2}$ (184 pm) is smaller than that of Se$^{-2}$ (191 pm), indicating that S$^{-2}$ can uniformly substitute for Se$^{-2}$ in TlNi$_2$Se$_{2-x}$S$_x$ system. The lattice parameters $\textit{a}$ = 3.87${\AA}$, $\textit{c}$ = 13.43${\AA}$ for TlNi$_{2}$Se$_{2}$, and $\textit{a}$ = 3.79${\AA}$, $\textit{c}$ = 12.77${\AA}$ for the other end member TlNi$_{2}$S$_{2}$ and for other TlNi$_{2}$Se$_{2-x}$S$_x$ compounds (see the right inset of Fig.1) were obtained by the fitting of their XRD data. The measurements of electrical resistivity, and specific heat were carried out using the Quantum Design PPMS-9. The dc magnetic susceptibility measurements were done using Quantum Design MPMS-SQUID.

\begin{figure}
  \includegraphics[width=8cm]{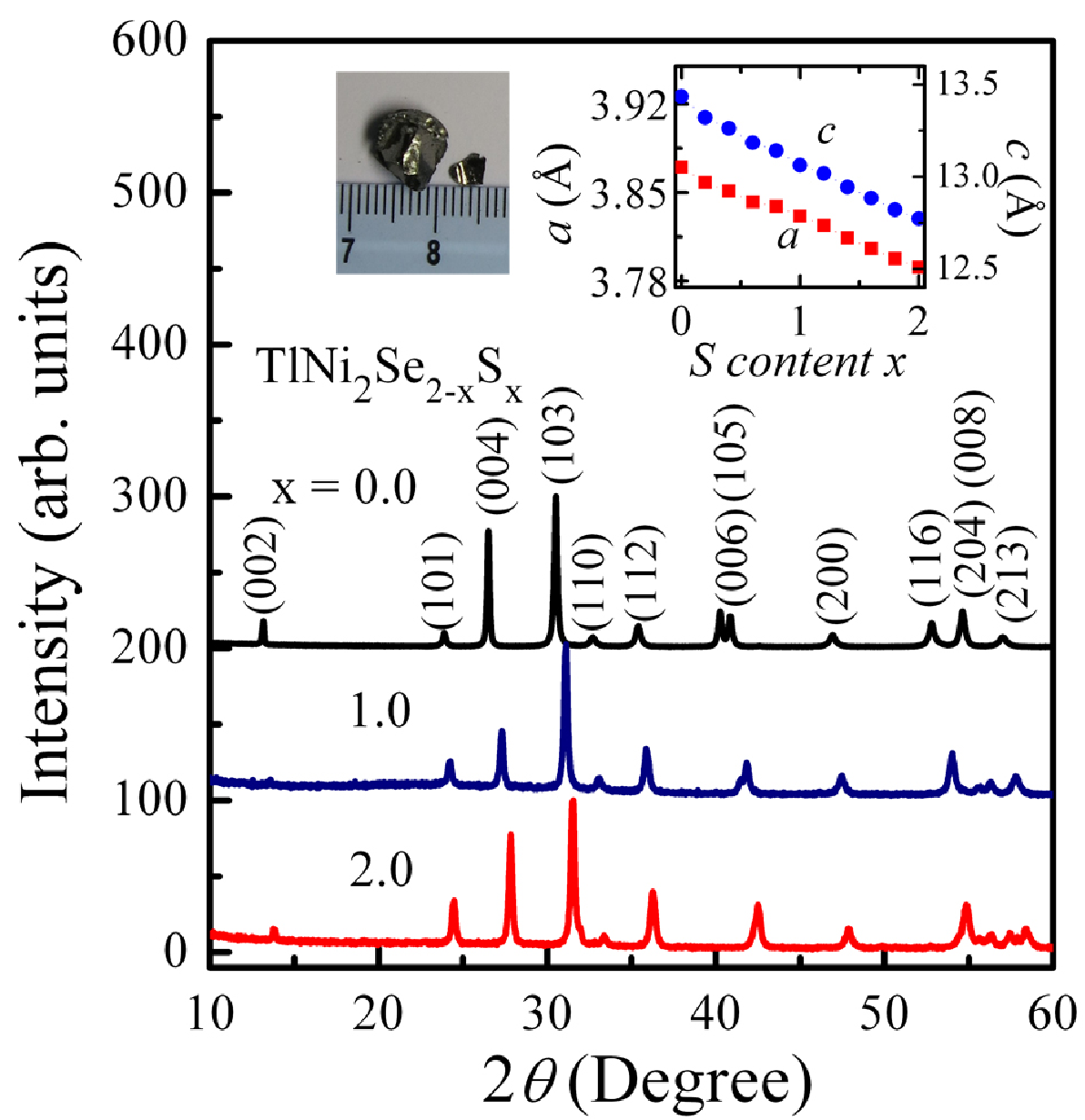}\\
  \caption{(Color online) X-ray diffraction pattern of powder obtained by grinding TlNi$_2$Se$_{2-x}$S$_x$ crystals (x=0.0, 1.0 and 2.0). Left inset: Photo of TlNi$2$S$_2$ crystal; right inset: the S content, \textit{x}, dependence of the lattice parameter \textit{a} and \textit{c}.}\label{}
\end{figure}

\begin{figure}
\includegraphics[width=8cm]{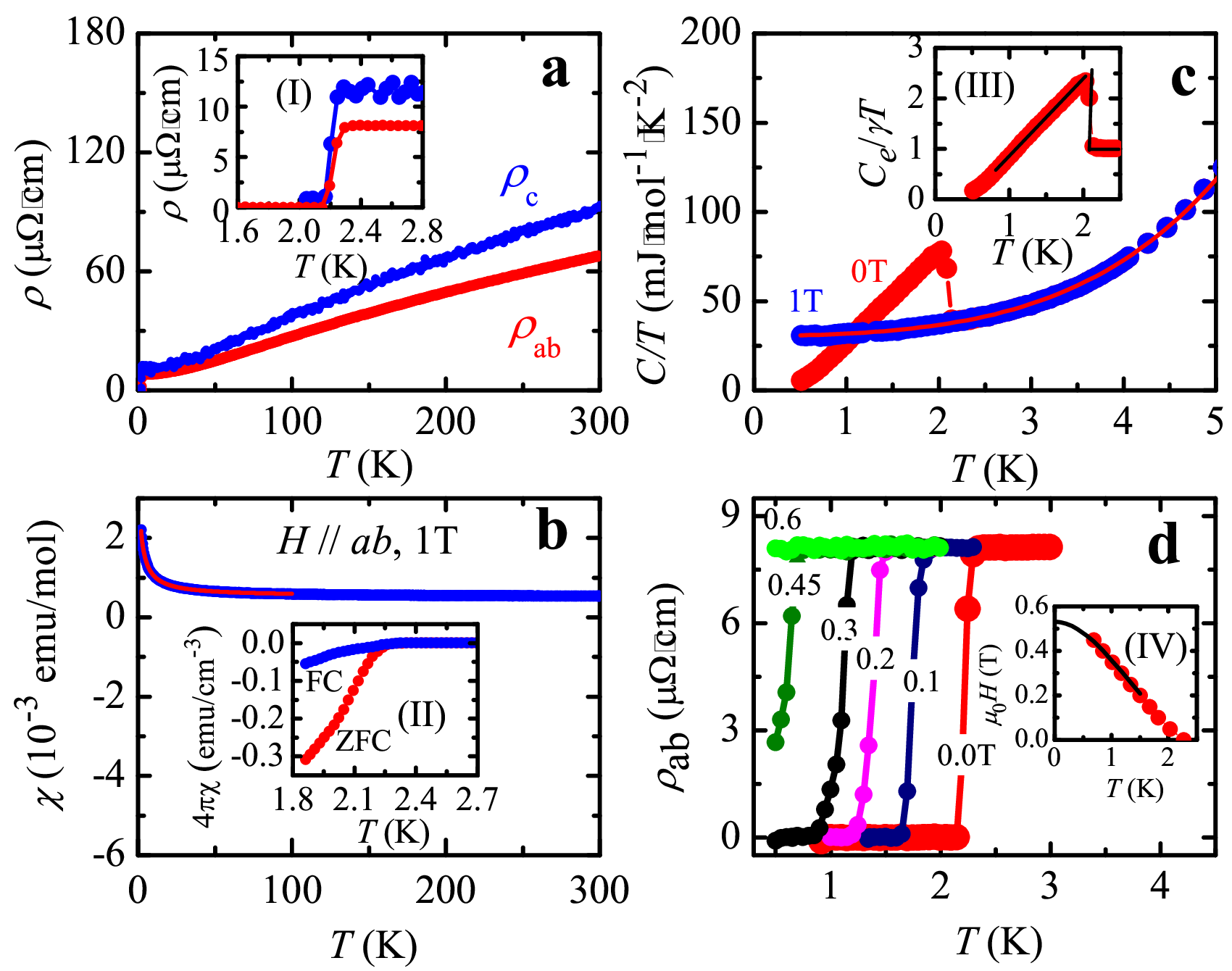}\\
\caption{(Color online)(a) Temperature dependence of both \textit{in}  and \textit{out-of}-plane resistivity, $\rho$$_{ab}$(T) and $\rho$$_c$(T) for TlNi$_2$S$_2$ single crystal. (b) Temperature dependence of the normal state magnetic susceptibility, $\chi$(T), measured at 1 Tesla field parallel to \textit{ab} plane. (c) Specific heat divided by temperature, $C/T$ vs \textit{T}, measured under 0 and 1Tesla field. (d) $\rho$$_{ab}$(T) data measured at different magnetic field near the superconducting transition. Inset: (i) $\rho$$_{ab}$(T) and $\rho$$_c$(T) near the superconducting transition, (ii) $\chi$(T) near the superconducting transition, measured at 5 Oe field parallel to \textit{ab} plane (for minimizing the demagnetization factor) with both zero-field cooling (ZFC) and field cooling (FC) processes, (iii) Temperature dependence of zero-field electronic specific heat in the superconducting state divided by temperature, $C_{es}$/$T$. (iv) Temperature dependence of upper critical field, $H_{C2}$(T). }\label{}
\end{figure}

The physical properties of TlNi$_{2}$S$_{2}$ are summarized in Fig. 2. Both $\rho_{ab}$ and $\rho_c$ vs \textit{T} curves, shown in Fig. 2(a) and inset (i), display a metallic behavior in the normal state before dropping abruptly to zero when superconductivity occurs at $T_C$=2.3 K, which is also confirmed by a large diamagnetic signal [see the inset (ii) of Fig. 2] and a specific heat jump at $T_C$ as shown in Fig. 2(c). At first, we discuss the resistivity in the normal state. $\rho_{ab}$ and  $\rho_c$ at 300K is of 68 and 92 $\mu\Omega$$\cdot$$cm$, respectively, and $\rho_c$/$\rho_{ab}$= 1.35, indicating that the anisotropy in TlNi$_2$S$_2$ is rather small, although the compound has a layer structure. In the normal state, no abnormal behavior related to structural or magnetic transition is observed in the resistivity curves, as it usually happens in the iron-based superconductors, and the other iso-structural compounds, such as BaNi$_{2}$As$_{2}$ \cite{Ronning2008} and SrNi$_{2}$P$_{2}$ \cite{Ronning2009}. The temperature dependence of magnetic susceptibility, $\chi$ (T), in the normal state was measured at 1T field, as shown in Fig. 2(b). At higher temperatures, $\chi$ (T) shows almost temperature-independent Pauli paramagnetic behavior. At lower temperatures, an obvious Curie tail occurs, which may likely come from the magnetic impurity. The fit by Curie-Weiss law below 100K yields a temperature independent contribution $\chi_{0}$=5.3$\times$10$^{-4}$ emu/mol and Curie constant $\textit{C}$ = 0.0063, corresponding to 0.63 mol$\%$ of an $S$ = 1 impurity ($\textit{e.g.}$, Ni$^{2+}$), indicating the absence of localized magnetism in TlNi$_2$S$_2$ crystal, similar to that in TlNi$_2$Se$_2$.

Figure 2(c) shows the temperature dependence of the specific heat divided by temperature, $C(T)$/$T$, measured at both zero field and 1 Tesla field. At zero field, $C(T)$/$T$ curve reveals a clear $\lambda$ anomaly, indicating the bulk nature of superconductivity. The normal state specific heat $C_{N}$(T) is obtained by applying a magnetic field $\mu_{0}H$ = 1.0 T. The \textit{C/T} $\textit{vs }$. $T^{2}$ curve (not shown here) shows nonlinear behavior at low temperatures, so we fitted the data using a polynomial expansion C/T = $\gamma_{N}$ + $\beta$T$^{2}$ + $\delta$T$^{4}$ below 6K. $\gamma_{N}$ = 30.57 mJ/mol$\cdot$K$^{2}$, and $\beta$ = 1.14 mJ/mol$\cdot$K$^{4}$ were obtained, corresponding to a Debye temperature $\Theta$ = 197K. The value of normal state electronic specific heat coefficient $\gamma_{N}$ is smaller than that of TlNi$_{2}$Se$_{2}$ ($\sim$40mJ/mol$\cdot$K$^{2}$) and KNi$_{2}$Se$_{2}$ ($\sim$44mJ/mol$\cdot$K$^{2}$) superconductors. However, it is still much larger than that of LaO$_{1-x}$F$_{x}$NiAs ($\sim$7.3mJ/mol$\cdot$K$^{2}$) \cite{Li 2008} and BaNi$_{2}$As$_{2}$ ($\sim$12.3mJ/mol$\cdot$K$^{2}$) \cite{Ronning2008}. Assuming 1.5 carriers/Ni and a spherical fermi surface, it yields m$^{*} \sim$ 13m$_{e}$. The zero-field electronic specific heat in the superconducting state, $C_{es}$, was obtained by subtracting the phonon contribution estimated by the $C$(T) measured at 1 T. The inset (iii) in Fig.2 shows the temperature dependence of C$_{es}$/$\gamma_{N}$T. The specific heat jump at $T_C$, $\Delta$C/$\gamma_{N}$T$_{C}$ = 1.55, is slightly larger than the weak-coupling BCS expectation of 1.43. Then, we use two different formula, based on the conventional BCS theory and two-gap model, respectively, to fit the data of C$_{es}$(T) below \textit{T}$_{C}$. Though the measured temperature is not low enough to get the best parameters, it still can be obtained that the BCS model cannot well described the C$_{es}$(T) data and the two-gap model presents the best fit (not shown here). The yielded two superconducting gaps are $\Delta$$_{1}$/k$_{B}$T$_{C}$=0.95 and $\Delta$$_{2}$/k$_{B}$T$_{C}$=1.96, respectively, and the relative weight contributed from the first gap is about 0.18. This result is very similar with that of TlNi$_2$Se$_2$, indicating that they may contain the similar electronic structure \cite{Goh 2014}.

The temperature dependence of upper critical field, $H_{C2}$(T), obtained from $\rho_{ab}$(T) measurements at various magnetic field, is shown in the inset (iv) of Fig. 2. It is found that $H_{C2}$ can be well described by the Ginzburg-Landau model, $H_{C2}$(T) = $H_{C2}$(0)$\times$(1-t$^{2}$)/(1+t$^{2}$), where \textit{t} is the reduced temperature $T/T_{C}$. Then the zero temperature upper critical field $H_{C2}$(0) is estimated to be about 0.53T. The superconducting coherence length $\xi_0$ can be estimated to be 24.9 nm from the relation $\xi_0$=[$\Phi_0$/2$\pi$$H_{C2}$]$^{1/2}$. Then the Fermi velocity $v_F$=4.09$\times$10$^4$ m/s is obtained from $\xi_0$=0.18$\hbar$$v_F$/$k_B$$T_C$. Using a spherical Fermi surface approximation, the Fermi wave vector is given by $k_F$=(3$\pi^2$Z/$\Omega$)$^{1/3}$, where \textit{Z} is the number of electrons per unit cell and $\Omega$ is the unit cell volume. Assuming that Ni contributes 1.5 electrons (\textit{Z}=6), we obtain $k_F$=9.89$\times$10$^9$ m$^{-1}$. Then m* and $\gamma_N$ can be estimated using the expression $m^*$=$\hbar$$k_F$/$v_F$ and $\gamma_N$=$\pi^2$$N$$k_B^2$$m^*$/$\hbar^2$$k_F^2$, which yields m*$\sim$ 25.5 m$_e$ and $\gamma_N$ $\sim$ 73.3 mJ/mol K$^2$, respectively. The values of $m^*$ and $\gamma_N$ are comparable to the values estimated from the normal state specific heat.

\begin{figure}
  \includegraphics[width=6cm]{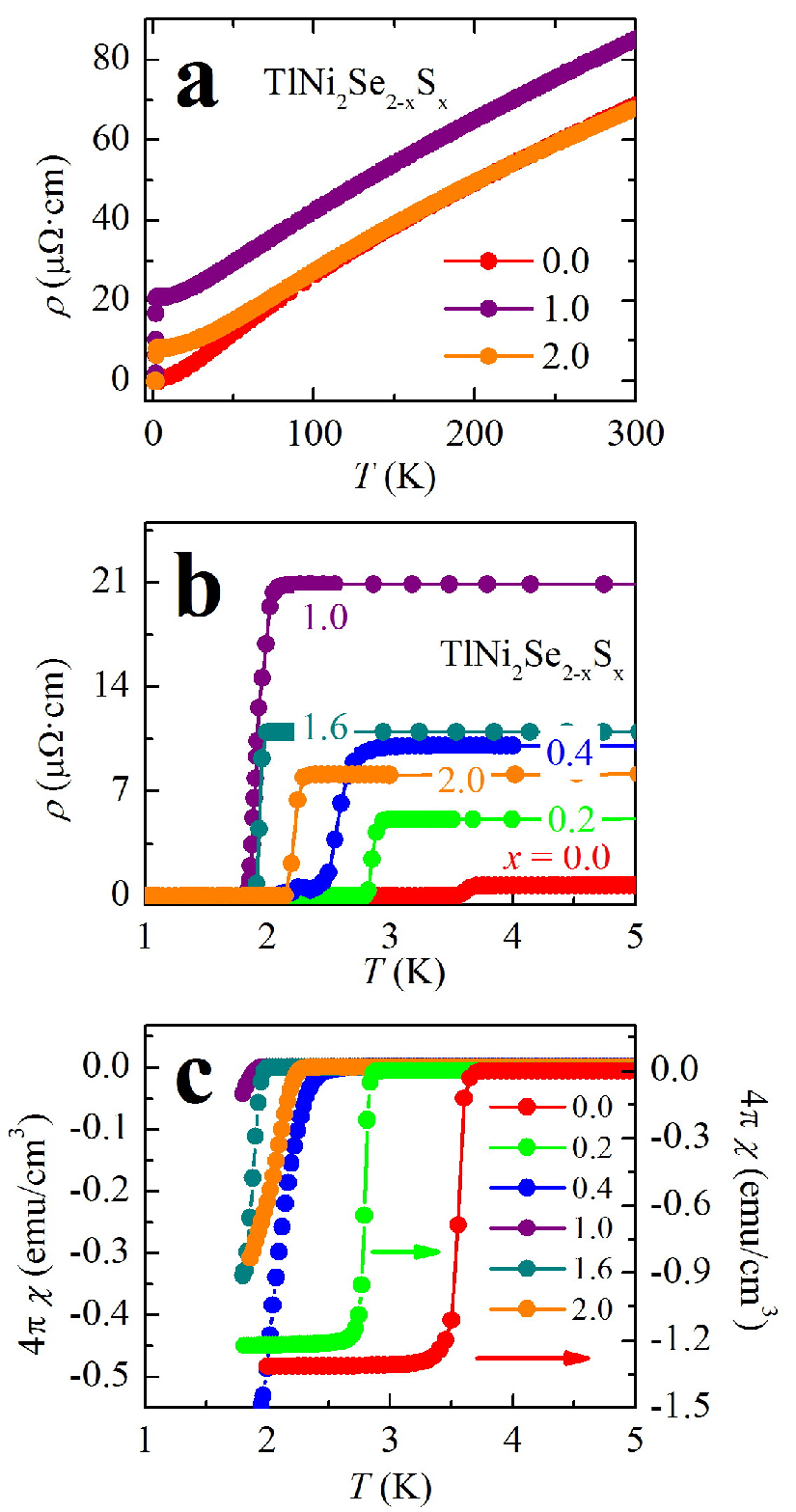}\\
  \caption{(Color online) (a) Temperature dependence of resistivity for TlNi$_2$Se$_{2-x}$S$_x$ single crystals. (b) The magnified view of the $\rho$(T) near the superconducting transition. (c) Temperature dependence of susceptibility, $\chi$(T), near the superconducting transition, measured at 5 Oe field with ZFC process.}\label{}
\end{figure}

The effect of S substitution for Se on superconductivity were systematically studied for TlNi$_2$Se$_{2-x}$S$_x$ system. Fig. 3a is the temperature dependence of resistivity for TlNi$_{2}$Se$_{2-x}$S$_{x}$ with $\textit{x}$ = 0.0, 1.0 and 2.0 as typical examples. The measurement reveals that all the samples exhibit metallic behavior at normal state, and undergo a superconducting transition at low temperatures. The superconducting transition is confirmed both by resistivity and magnetic measurement. The magnified view of the temperature dependence of resistivity and susceptibility at low temperatures of typical compositions in the TlNi$_{2}$Se$_{2-x}$S$_{x}$ series are summarized in Fig. 3b and 3c, respectively. As shown in the figure, the transition temperature, yielded from resistivity and susceptibility analysis, respectively, are consistent with each other.

To get more information about the superconducting properties, we also carried out specific heat measurement in several typical TlNi$_2$Se$_{2-x}$S$_x$ samples with $\textit{x}$ = 0.2, 0.8, 1.2, and 2.0. The results are presented in Fig. 4. At zero field, a superconducting transition is clearly observed at low temperatures for all the samples, indicating the bulk nature of superconductivity. The normal state specific heat \textit{C}$_{N}$(T) was estimated by applying a 1T magnetic field. In all the samples, no Schottky anomalies were detected. Using the polynomial expansion C/T = $\gamma_{N}$ + $\beta$T$^{2}$ + $\delta$T$^{4}$ to fit the data below 6K, we got the normal state parameters $\gamma$$_{N}$, $\beta$, and $\delta$. The results are summarized in table I. With increasing of the S content, $\textit{x}$, the electronic specific heat coefficient $\gamma$$_{N}$ decreases from 40mJ/mol$\cdot$K$^{2}$ in TlNi$_2$Se$_{2}$ to $\sim$35.6mJ/mol$\cdot$K$^{2}$ in the sample with $\textit{x}$ = 0.2, and further goes to $\sim$30mJ/mol$\cdot$K$^{2}$ in the samples with $\textit{x}$ = 0.8, 1.2 and 2.0. However, they are still much larger than that of the other nickel-based superconductors, indicating that the heavy electron property preserves in all the TlNi$_2$Se$_{2-x}$S$_x$ samples. Here, we also carried out the fitting using the two-gap model for the $C_{es}$(T) data, which can be yielded by subtracting the phonon contribution estimated by the $C$(T) measured at 1 T. However, the fitting results are not good, since the measured temperature is not low enough.

\begin{figure}
  \includegraphics[width=8cm]{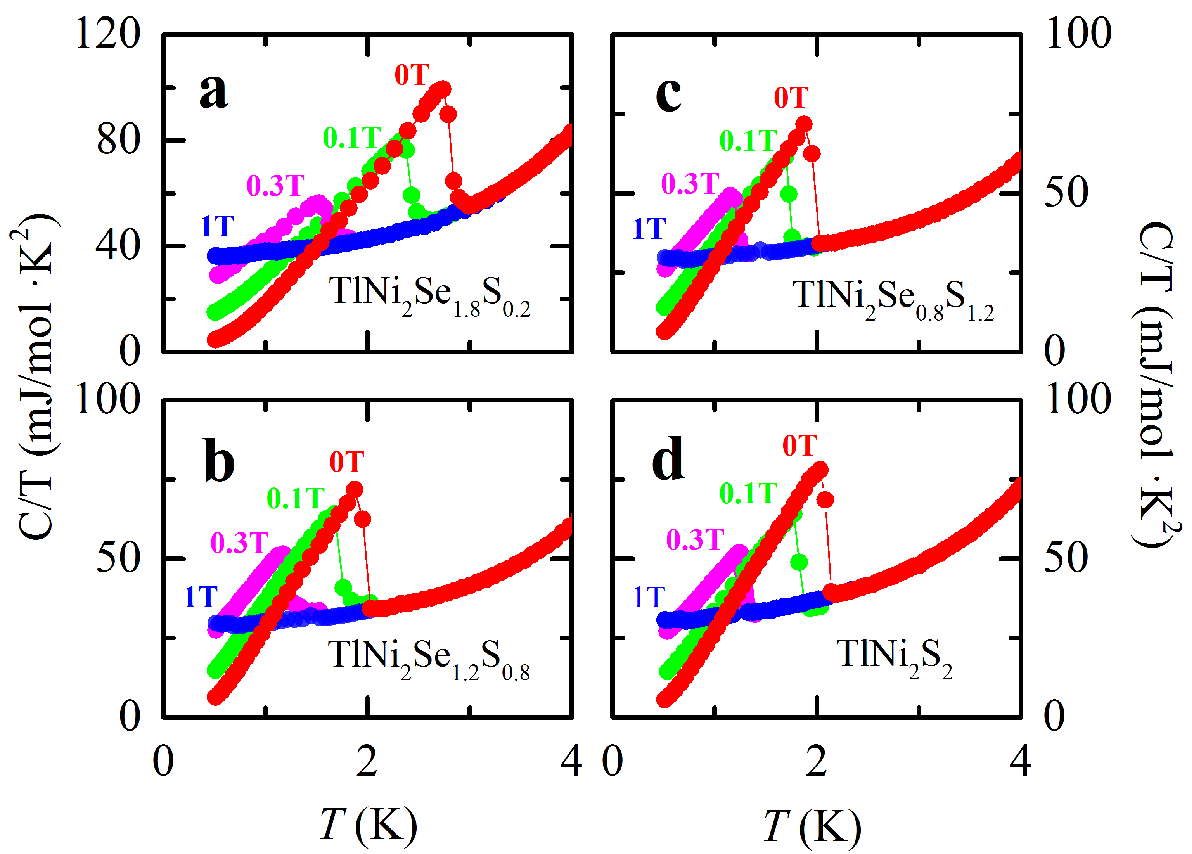}\\
  \caption{Low temperature specific heat divided by temperature, $\textit{C/T}$ vs. T, measured at variant magnetic field, for (a) TlNi$_2$Se$_{1.8}$S$_{0.2}$, (b) TlNi$_2$Se$_{1.2}$S$_{0.8}$, (c) TlNi$_2$Se$_{0.8}$S$_{1.2}$ and (d) TlNiS$_2$, respectively. To make the figure clear, not all the datas are presented.}\label{}
\end{figure}

\begin{table}
  \centering
  \caption{The fitting results of the specific heat for TlNi$_2$Se$_{2-x}$S$_{x}$ crystals. (The data for TlNiSe$_2$ comes from ref. \cite{WangHD2013})}
\begin{tabular}{l|c|c|c}
  \hline \hline
    & $\gamma$ & $\beta$ & $\delta$ \\
    & (mJ/mol$\cdot$K$^{2}$) & (mJ/mol$\cdot$K$^{4}$) & (mJ/mol$\cdot$K$^{6}$)\\\hline
  TlNiSe$_2$ & 40 & 1.65 & 0.135 \\
  TlNi$_2$Se$_{1.8}$S$_{0.2}$ & 35.68 & 1.32 & 0.104\\
  TlNi$_2$Se$_{1.2}$S$_{0.8}$ & 31.13 & 0.744 & 0.093\\
  TlNi$_2$Se$_{0.8}$S$_{1.2}$ & 29.66 & 0.511 & 0.090\\
  TlNiS$_2$ & 30.57 & 1.14 & 0.094\\
  \hline \hline
\end{tabular}
\end{table}

\begin{figure}
  \includegraphics[width=6cm]{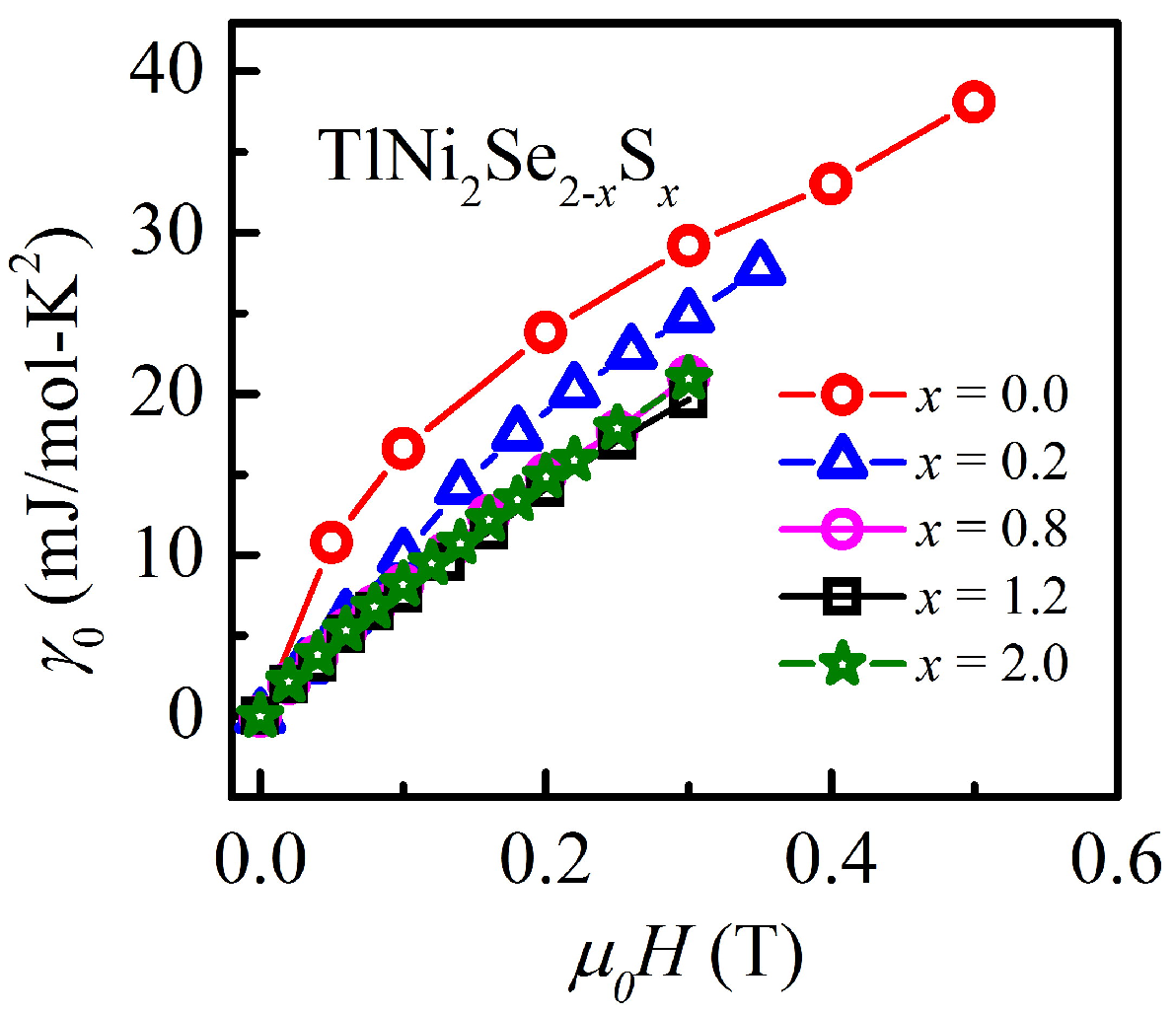}\\
  \caption{Magnetic field $\mu$$_{0}$H dependence of electronic specific heat coefficient $\gamma$$_{0}$ in the mixed state for TlNi$_2$Se$_{2-x}$S$_x$ with $\textit{x}$=0.0, 0.2, 0.8, 1.2 and 2.0, respectively.}\label{}
\end{figure}

In our previous article \cite{WangHD2013}, we reported a H$^{1/2}$ relation for the magnetic field dependence of the residual specific heat coefficient $\gamma$$_{N}$ at T=0K in the mixed state in TlNi$_2$Se$_{2}$, which is usually considered as a common feature of the d-wave superconductors. To figure out how the sulfur doping affects the behavior, we measured the specific heat of TlNi$_2$Se$_{2-x}$S$_x$ at various magnetic field. The results are shown in Fig. 4. With increasing of field, the superconducting transition is suppressed. Using the method described in ref. \cite{WangHD2013}, we got the residual specific heat coefficient $\gamma$$_{N}$ in the mixed state. The field dependence of $\gamma$$_{N}$ for TlNi$_2$Se$_{2-x}$S$_x$ with $\textit{x}$ = 0.0, 0.2, 0.8, 1.2 and 2.0 are plotted in Fig. 5. From the specific heat data, the $\gamma$$_{N}$ for TlNi$_2$Se$_{2}$ exhibits a convex variation with field (proportional to H$^{1/2}$). With the S starting to substitute Se, $\textit{i.e.}$, for TlNi$_2$Se$_{1.8}$S$_{0.2}$, the $\gamma$$_{N}$-$\textit{H}$ curve becomes not so convex. Then, with further increasing of the S content, $\textit{i.e.}$, for TlNi$_2$Se$_{2-x}$S$_x$ with $\textit{x}$ = 0.8, 1.2 and 2.0, the $\gamma$$_{N}$ even exhibits a linear behavior, which is consistent with that of conventional s-wave superconductors. The variation of the $\gamma$$_{N}$-$\textit{H}$ curves is very interesting and exotic, since both the TlNi$_2$Se$_{2}$ and TlNi$_2$S$_2$ exhibit the similar physical properties in the normal state. In authors' view \cite{WangHD2013}, a convex behavior for the field dependence of $\gamma$$_{N}$ may related to the strength of the vortex-vortex interactions in the mixed state \cite{Ramirez1996}, and not be responding to the exotic pairing state. Besides, such behavior has also been detected in some iron-based superconductors \cite{Wen2009,Wen2010,Wen2008}. Bang \cite{Bang2010} has proposed that this phenomena can be resulted by the variation of the superconducting gap size ratio, $\Delta$$_{1}$/$\Delta$$_{2}$, and is a generic feature of the multi-band superconductors. However, further investigations, like STM and ARPES, are needed to determine the superconducting order parameter symmetry. Recently, low-temperature thermal conductivity measurements revealed the multiple nodeless superconducting gaps in TlNi$_2$Se$_{2}$ compound \cite{Hong2014}, which is consistent with our opinion. From this point view, the superconducting properties in TlNi$_2$Se$_{2}$ and TlNi$_2$S$_2$ may not differ with each other so much.

To clarify the superconducting properties, the critical temperature \textit{T}$_{C}$ were depicted in Fig. 6 as a function of S content $\textit{x}$. The critical temperature is determined by resistivity, susceptibility, and specific heat measurement, respectively. A very small difference is detected for these values, which may contributed to the inhomogeneous of the samples. Here we use the \textit{T}$_{C}$ value yielded from $\rho$(T) to keep consistent. For the undoped sample, $\textit{i.e.}$ TlNi$_{2}$Se$_{2}$, the \textit{T}$_{C}$ is 3.7K, much higher than that of KNi$_{2}$Se$_{2}$ ($\sim$0.8K) and KNi$_{2}$S$_{2}$ ($\sim$0.4K). With the increase of S content $\textit{x}$, the \textit{T}$_{C}$ declines to be about 2K for $\textit{x}$ = 0.6, and changes a little for the samples with 0.6 $\leq$ $\textit{x}$ $\leq$ 1.6. With the S content $\textit{x}$ further increasing, the \textit{T}$_{C}$ has a little increase and achieves to be $\sim$2.25K for the end member TlNi$_{2}$S$_{2}$.

\begin{figure}
  \includegraphics[width=8cm]{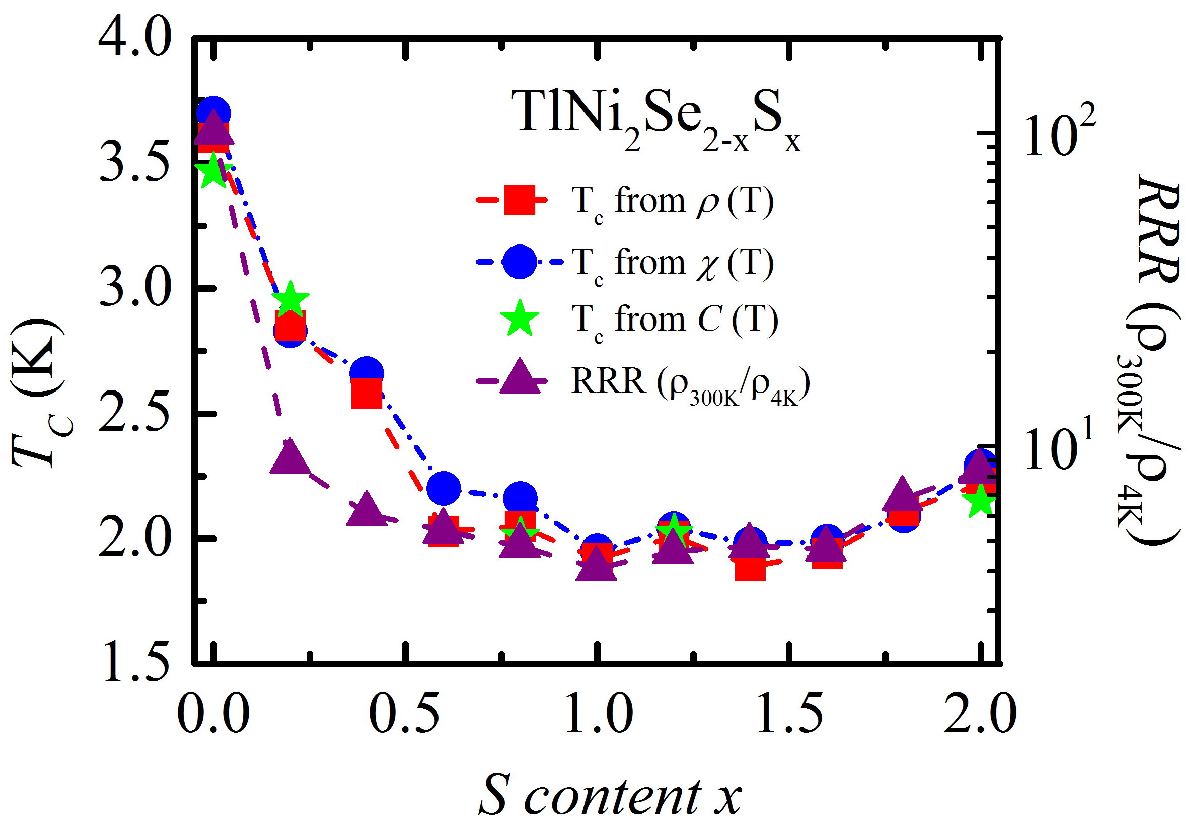}\\
  \caption{(Color online) The critical temperature $\textit{T}$$_{C}$ and the residual resistivity ratio (RRR) as a function of S content $\textit{x}$ in TlNi$_2$Se$_{2-x}$S$_x$. }\label{}
\end{figure}

In TlNi$_{2}$Se$_{2-x}$S$_{x}$ system, the partially substitution of Selenium with Sulfur doesn't introduce additional electrons or holes. From the structural point view, the S substitution effect is equivalent to apply a hydrostatic pressure. Usually, it has a prominent effect on the superconducting transition temperature. So the change of the critical temperature \textit{T}$_{C}$ with the S content $\textit{x}$ is somehow beyond our expectation. To understand the S substitution effect on the superconducting properties, we reexamined the transport properties of TlNi$_{2}$Se$_{2-x}$S$_{x}$. For the undoped sample TlNi$_{2}$Se$_{2}$, the residual resistivity ratio (RRR), which is defined as $\rho$$_{300K}$/$\rho$$_{4K}$, is calculated to be about 100. For the doped samples, though the value of resistivity changes slightly with the varying of $\textit{S}$ content $\textit{x}$ at room temperature (T=300K), the value of resistivity at low temperatures is about one order magnitude larger than that of TlNi$_{2}$Se$_{2}$, resulting much smaller RRR values. Especially for TlNi$_{2}$SeS, where half Se sites are replaced by S, the temperature dependence of resistivity exhibits a more localized behavior at normal state. We suggest that this pronounced phenomena may be contributed to the disorder introduced by S substitution in TlNi$_{2}$Se$_{2-x}$S$_{x}$ series. To investigate the relationship between the critical temperature \textit{T}$_{C}$ and the disorder effect, we depict the RRR value as a function of S content $\textit{x}$ in the Fig.6. As shown in the figure, TlNi$_{2}$Se$_{2}$ shows the largest RRR value. With the increase of $\textit{S}$ content $\textit{x}$, the RRR value decreases drastically, and achieves the smallest value ($\sim$4.1) for the sample with $\textit{x}$ = 1.0. With S content further increasing, the RRR value increases slightly, and achieves 8.3 for TlNi$_{2}$S$_{2}$. The variation of the RRR value is similar with that of \textit{T}$_{C}$ as discussed before, suggesting that the disorder effect is playing an important role in the superconducting transition temperature in this system.

In the end, we synthesized successfully a series of TlNi$_2$Se$_{2-x}$S$_x$ (0.0 $\leq$ x $\leq$ 2.0) single crystals. The superconductivity with \textit{T}$_{C}$$\sim$ 2.3K was first detected in TlNi$_2$S$_{2}$ crystal, which appears to involve heavy electrons with an effective mass m*=13$\thicksim$25 m$_{e}$. With the S substitution, the superconductivity and heavy electron behavior was preserved in all the TlNi$_2$Se$_{2-x}$S$_x$ samples. However, in the mixed state, a novel change of the field dependence of the residual specific heat coefficient, $\gamma$$_{N}$(H), occurs in TlNi$_2$Se$_{2-x}$S$_x$, with increasing of the S content, $\textit{x}$. On the other hand, it was found that the \textit{T}$_C$ value in the TlNi$_2$Se$_{2-x}$S$_x$ crystals is related to the disorder, characterized by the RRR, which is induced by the partial substitution of S for Se. We suggest that TlNi$_2$Se$_{2-x}$S$_x$ system provides an example to study the effect of disorder on the multi-band superconductivity.

This work is supported by the National Basic Research Program of China (973 Program) under grant No. 2011CBA00103, 2012CB821404 and 2015CB921004, the Nature Science Foundation of China (Grant No. 11374261, and 11204059) and Zhejiang Provincial Natural Science Foundation of China (Grant No. LQ12A04007), and the Fundamental Research Funds for the Central Universities of China.


\begin{thebibliography}{apssamp}
\bibitem{ChenXH2009} X. F. Wang, T. Wu, G. Wu, H. Chen, Y. L. Xie, J. J. Ying, Y. J. Yan, R. H. Liu, and X. H. Chen, Phys. Rev. Lett. \textbf{102}, 117005(2009)

\bibitem{YangJH2013} J. H. Yang, B. Chen, H. D. Wang, Q. H. Mao, M. Imai, K. Yoshimura, and M. H. Fang, Phys. Rev. B \textbf{88}, 064406(2013)

\bibitem{Rotter2008} M. Rotter, M. Tegel, and D. Johrendt, Phys. Rev. Lett. \textbf{101}, 107006(2008)

\bibitem{Fang2011} M. H. Fang, H. D. Wang, C. H. Dong, Z. J. Li, C. M. Feng, J. Chen, and H. Q. Yuan, EPL \textbf{94}, 27009(2011)

\bibitem{Wang2011} H. D. Wang, C. H. Dong, Z. J. Li, Q. H. Mao, S. S. Zhu, C. M. Feng, H. Q. Yuan, and M. H. Fang,, EPL \textbf{93}, 47004(2011)

\bibitem{Steglich1979} F. Steglich, J. Aarts, C. D. Bredl, W. Lieke, D. Meschede, W. Franz, and H. Sch$\ddot{a}$fer, Phys. Rev. Lett. \textbf{43}, 1892(1979)

\bibitem{Ronning2008} F. Ronning, N. Kurita, E. D. Bauer, B. L. Scott, T. Park, T. Klimczuk, R. Movshovich, and J. D. Thompson, J. Phys. Condens. Matter \textbf{20}, 342203(2008)

\bibitem{Ronning2009} F. Ronning, E. D. Bauer, T. Park, S. H. Baek, H. Sakai, and J. D. Thompson, Phy. Rev. B. \textbf{79}, 134507(2009)

\bibitem{Neilson2012-1} J. R. Neilson, Anna Llobet, A.V. Stier, L. Wu, J. Wen, J. Tao, Y. Zhu, Z. B. Tesanovic, N. P. Armitage, and T. M. McQueen, Phys. Rev. B \textbf{86}, 054512(2012)

\bibitem{Neilson2013} J. R. Neilson, T. M. McQueen, A. Llobet, J. J. Wen, and M. R. Suchomel, Phys. Rev. B \textbf{87}, 045127(2013)

\bibitem{WangHD2013} H. D. Wang, C. H. Dong, Q. H. Mao, Rajwali Khan, X. Zhou, C. X. Li, B. Chen, J. H. Yang, Q. P. Su, and M. H. Fang, Phys. Rev. Lett. \textbf{111}, 207001(2013)

\bibitem{Murray2012} James M. Murray and Zlatko Te$\check{s}$anovi$\acute{c}$, Phys. Rev. B \textbf{87}, 081103(2013)

\bibitem{Newmark 1989} A.R. Newmark, G. Huan, M. Greenblatt, M. Croft, Solid State Commun., \textbf{71}, 1025(1989)

\bibitem{Li 2008} Z. Li, G. F. Chen, J. Dong, G. Li, W. Z. Hu, D. Wu, S. K. Su, P. Zheng, T. Xiang, N. L. Wang, J. L. Luo, Phys. Rev. B. \textbf{78}, 060504(R)(2008)

\bibitem{Goh 2014} S. K. Goh, H. C. Chang, P. Reiss, P. L. Alireza, Y. W. Cheung, S. Y. Lau, Hangdong Wang, Qianhui Mao, Jinhu Yang, Minghu Fang, F. M. Grosche, and M. L. Sutherland, Phys. Rev. B $\textbf{90}$, 201105(R) (2014)

\bibitem{Ramirez1996} A. P. Ramirez, Phys. Lett. A \textbf{211}, 59(1996)

\bibitem{Wen2009} G. Mu, H. Luo, Z. Wang, L. Shan, C. Ren, and H. H. Wen, Phys. Rev. B \textbf{79}, 174501 (2009)

\bibitem{Wen2010} G. Mu, B. Zeng, P. Cheng, Z. S. Wang, L. Fang, B. Shen, L. Shan, C. Ren and H. H. Wen, Chin. Phys. Lett. \textbf{27}, 037402 (2010)

\bibitem{Wen2008} G. Mu, X. Y. Zhu, L. Fang, L. Shan, C. Ren and H. H. Wen, \textbf{25}, 2221 (2008)

\bibitem{Bang2010} Y. Bang, Phys. Rev. Lett. \textbf{104}, 217001 (2010)

\bibitem{Hong2014} X. C. Hong, Z. Zhang, S. Y. Zhou, J. Pan, Y. Xu, H. D. Wang, Q. H. Mao, M. H. Fang, J. K. Dong, and S. Y. Li, Phys. Rev. B \textbf{90}, 060504(R)(2014)

\end{thebibliography}
\end{document}